\documentclass[aps,pra,reprint,superscriptaddress,longbibliography,footinbib,floatfix]{revtex4-2}
\usepackage[colorlinks=true, urlcolor=teal,linkcolor=teal, citecolor=teal]{hyperref}
\usepackage[english]{babel}
\usepackage[dvipsnames]{xcolor}
\usepackage[T1]{fontenc}
\usepackage[utf8]{inputenc} 
\usepackage{ragged2e}
\usepackage{graphicx}
\usepackage{array}
\usepackage{booktabs}
\usepackage{amsmath,amsfonts,amsthm}
\usepackage{caption}
\usepackage{mathtools}

\begingroup

\footnotetext{J.M. and S.P.M. contributed equally to this work.}
\endgroup

\begin{document}

\title{Bath memory as a precision resource in quantum transport}

\author{José Molina}
\email{j.molinacano@um.es}
\author{Sheikh Parvez Mandal}
\author{Mahasweta Pandit}
\email{m.pandit@um.es}
\author{Javier Prior}
\email{javier.prior@um.es}
\affiliation{Departamento de Física - CIOyN, Universidad de Murcia, Murcia E-30071, Spain}

\begin{abstract}
Structured baths can reshape transport fluctuations in mesoscopic quantum devices, yet a predictive criterion for when this enhances precision has been lacking. We propose a route towards such precision advantages by utilizing bath memory in coherent fermionic transport through a noninteracting quantum-dot chain. Using the Landauer--B\"uttiker formalism, we derive a dual impedance-matching condition that synchronizes the conductor mode splitting, boundary dissipation, and bath bandwidth, and sustains constructive multimode interference across the transmission window. The analytical predictions for the optimal bath bandwidths show excellent agreement with exact nonequilibrium Green's function calculations of the transport for Lorentzian, Gaussian, and Newns spectral densities. The prescription yields an optimal bath bandwidth at which the current Fano factor is minimized and the thermodynamic and kinetic precision coefficients are simultaneously enhanced beyond their Markovian limits. The alignment of the optimal precision regime with the experimentally accessible current Fano factor minimum thus provides a practical strategy for designing precision-enhanced transport in mesoscopic platforms such as semiconductor quantum-dot arrays and ultracold fermionic channels.
\end{abstract}

\maketitle

\label{sec:outline}

Mesoscopic transport provides a powerful framework for understanding how currents flow in nanoscale quantum devices, where fluctuations, dissipation, and coherence interplay in a nontrivial manner~\cite{datta1997electronic, blanter2000shot,buttiker1992scattering}. Beyond its practical relevance for quantum electronics and energy conversion \cite{landauer1957, Landauer}, it also provides a natural bridge to quantum thermodynamics \cite{Skrzypczyk_2014, Info, Hasegawa2020, mandal2026heat}. In this setting, thermodynamic and kinetic uncertainty relations (TUR and KUR) not only constrain current precision by linking relevant quantities such as entropy production and dynamical activity~\cite{barato2015,gingrich2016,pietzonka2018,KUR,TKUR}, but also underpin inference schemes to bound these quantities using current fluctuations, waiting-time statistics, and partial observations~\cite{gingrich2017inferring, li2019quantifying, manikandan2020inferring, vu2020entropy, skinner2021estimating, vandermeer2022thermo, nitzan2023universal, vo2022unified}.

Quantum extensions of the thermodynamic and kinetic uncertainty relations valid for coherent quantum transport have been proposed~\cite{PhysRevLett.134.020401, markovian,brandner2025thermodynamic, blasi2026quantumkineticuncertaintyrelations, PhysRevE.104.L012103, PhysRevB.98.155438, PhysRevE.103.012133}. Yet, the effect of bath memory on current precision remains less settled and largely model-dependent. Finite bath correlation times have been reported to either hinder transport by creating memory-induced bottlenecks in driven qubits and quantum heat engines~\cite{PhysRevA.99.052119, Maity_2024}, or enhance precision in models of spin-chain energy transport~\cite{PhysRevA.101.012123}, yet none of these settings yield a transferable criterion for coherent steady-state mesoscopic transport connecting the bath's memory timescale and precision advantages through experimentally measurable transport characteristics.

In this work, we address this issue and establish an analytical criterion for exploiting bath memory for precision advantages in coherent quantum transport. We link the conductor's spectral structure, bath dissipation, and bath correlation time, enabling simultaneous improvement of thermodynamic and kinetic precision coefficients. This yields a closed scaling relation [Eq.~\eqref{eq:theequation}] for Lorentzian reservoirs, which are then extended to other structured baths [Eq.~\eqref{eq:scaling_ansatz}] with Gaussian and Newns-type spectral densities. These findings are experimentally relevant because the finite-memory precision optimum is accompanied by a measurable minimum in the current Fano factor, providing a noise-based indicator of the memory-matched regime without requiring reconstruction of the microscopic dynamical activity. An extended reservoir or pseudomode formalism \cite{nazir2019reaction,  Mark, prior2010efficient, PhysRevA.80.012104, pleasance2020generalized} can also be used for implementing baths with tunable memory, making our predictions directly testable in existing mesoscopic platforms. 

\begin{figure}[b]
    \centering
    \includegraphics[trim=0.5cm 0.5cm 0.5cm 0.26cm, width=.9\linewidth]{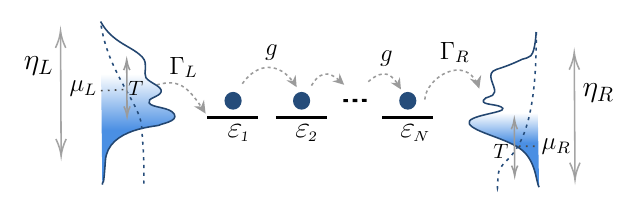}
    \parbox{0.5\textwidth}{
    \caption{\justifying Schematic representation of the quantum transport setup.\label{fig:schematic_firstpage}}}
    
\end{figure}

\textit{\textbf{Model.---}}We study nonequilibrium fermionic transport through a noninteracting one-dimensional quantum-dot array coupled to two fermionic baths at its left ($L$) and right ($R$) ends,
\begin{equation}
H_{\rm sys} = \sum_{i=1}^{N}\varepsilon_i c_i^\dagger c_{i}
+ \sum_{i=1}^{N-1} g_i \left(c_i^\dagger c_{i+1}+{\rm H.c.}\right),
\end{equation}
where $\varepsilon_i$ is the gate-tunable energy of the $i$th dot, and $g_i$ is the hopping between the $i$th and the $(i+1)$th dots. In this paper, we use a homogeneous chain, $g_i=g$ and $\epsilon_i=\epsilon$. The baths are described by Fermi functions $f_{\alpha}(\omega)
=\left[e^{\beta(\omega-\mu_{\alpha})}+1\right]^{-1}$, with $\alpha=L,R$, equal temperature $T=\beta^{-1}$, and symmetric-bias configuration $\mu_{L,R}=\epsilon\pm eV/2$ with chemical potentials $\mu_\alpha$ and voltage bias $eV$, unless otherwise stated. We use units $\hbar=k_B=e=1$ throughout the derivations. A schematic of the setup is shown in Fig.~\ref{fig:schematic_firstpage}. The model captures coherent transport relevant to ultracold fermionic channels~\cite{Brantut_2012, brantut2013thermoelectric, krinner2015observation, PhysRevX.8.011053} and semiconductor quantum-dot arrays~\cite{vanderWiel_2002, hensgens2017quantum, volk2019loading, mills2019shuttling}.

For the coherent noninteracting conductors considered here, the steady-state particle current can be obtained from the Landauer--B\"uttiker formalism~\cite{landauer1957, blanter2000shot,buttiker1992scattering, Landauer}, 
\begin{equation}
J = \frac{1}{2\pi} \int_{-\infty}^{\infty}d\omega\,
\mathcal{T}(\omega) \left[ f_L(\omega)-f_R(\omega) \right],
\label{eq:LB_current}
\end{equation}
where $\mathcal{T}(\omega)$ is the single-particle transmission probability. The zero-frequency noise can be calculated using the Levitov--Lesovik formula~\cite{levitov1993charge, schonhammer2007full}
\begin{multline}
    D = \frac{1}{2\pi}\int_{-\infty}^{\infty}d\omega\,\Big\{
\mathcal{T}(\omega) \left[ f_L(1-f_L)+f_R(1-f_R) \right]
\\ +\mathcal{T}(\omega) \left[ 1-\mathcal{T}(\omega) \right]
\left[ f_L-f_R \right]^2 \Big\},\label{eq:LB_noise_full}
\end{multline} 
where the frequency arguments of $f_{L,R}(\omega)$ are omitted for compactness. We define the nonequilibrium current Fano factor as $F:=D/J$.

Each fermionic bath is characterized by a spectral density $\mathcal{J}_\alpha(\omega)$. For the Lorentzian spectral density used for the analytical results, 
\begin{equation}
\mathcal{J}_\alpha(\omega) = \frac{\Gamma_\alpha (\eta_\alpha/2)^2}{(\omega-\varepsilon_{L_\alpha})^2+(\eta_\alpha/2)^2},
\label{eq:lorentzian_density_results}
\end{equation}
where $\varepsilon_{L_\alpha}$ is the frequency at which the bath coupling  reaches its maximum strength $\Gamma_\alpha$, while $\eta_\alpha$ sets the bath spectral bandwidth. A large $\eta_\alpha$ approaches the Markovian wide-band limit, whereas small $\eta_\alpha$ describes a narrow-band bath with longer correlation time. Unless stated otherwise, we focus on symmetric baths, $\Gamma_L=\Gamma_R=\Gamma$ and $\eta_L=\eta_R=\eta$. 
Lorentzian baths also admit a simple pseudomode representation~\cite{pleasance2020generalized,nazir2019reaction, Mark, pandit2026}. This allows the original $ N$-site chain coupled to Lorentzian baths to be represented by an extended Markovian system of $N+2$ quantum dots, with the energies of the auxiliary dots $\varepsilon_{L_\alpha}$, representing the left and right baths. We use this mapping to obtain analytical conditions for memory-assisted current noise suppression in these systems and give details in Appendix~\ref{ap:pseudo}.

\begin{figure*}[t]
    \centering
    \includegraphics[trim= 0.cm 1.cm .2cm 3.6cm, clip,width=\linewidth]{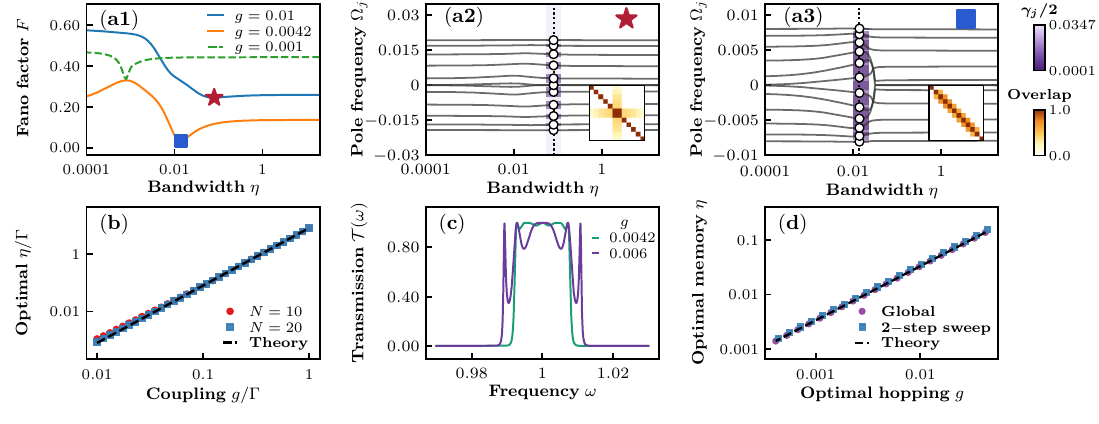}
    \parbox{\textwidth}{
    \caption{\justifying
    (a1) Current Fano factor $F$, calculated using the NEGF technique in the pseudomode representation of the Lorentzian baths, as a function of bath bandwidth $\eta$ for different hoppings $g$ with Fano minima marked for $g=0.01$ and $g=0.0042$.
    (a2, a3)  Pole frequencies $\Omega_j$ and damping rates $\gamma_j/2$ (color scale, right top) of the modes in Eq.~\eqref{eq:pole_expansion} for mismatched $(g=0.01)$ and optimal $g=0.0042$ internal hoppings as a function of bandwidth $\eta$. The circles represent the pole positions at the numerical optimal $\eta$. Insets show the inter-mode overlap matrix [Eq.~\eqref{eq:pole_expansion}] (color scale, right bottom). The optimal case ($g=0.0042$) displays strong uniform overlap, enabling constructive interference and noise suppression, while the mismatched case ($g=0.01$) shows suppressed off-diagonal overlaps. Noise suppression is observed in the optimal overlap case of $g=0.0042$ (a2).
    (b) Log-log plot of the numerical optimal pairs $(g,\eta)$ for $\Gamma=0.1$, compared with the theoretical prediction of Eq.~\eqref{eq:theequation}.
    (c) Transmission function for $N=5$ in both optimal $(g=0.0042)$ and overcoupled $(g=0.006)$ configurations, showing a flatter transmission profile for the former that suppresses the Fano factor.
    (d) Globally optimized and two-step pairs $(g,\eta)$ for $\Gamma \in [0.001,0.1]$ and the analytical prediction of Eq.~\eqref{eq:theequation}.
    Fixed parameters: $N=10$, $\Gamma_\alpha=0.01$, $T=1$, $eV/T=12$, and $\varepsilon_{L_\alpha}=\varepsilon=1$, unless stated otherwise.
    \label{Fig2}}} 
\end{figure*}

\textit{\textbf{Dual impedance-matching and current-noise suppression.---}}To study the mechanism analytically, we first consider the shot-noise-dominated transport limit in the high-bias regime, i.e., $eV \gg T$. We will show later that our results hold even when we relax this condition. In the high-bias regime, the Landauer--B\"uttiker expressions [Eqs.~\eqref{eq:LB_current}, \eqref{eq:LB_noise_full}] simplify because $f_L-f_R\simeq 1$ over the active transmission window and the equilibrium thermal contribution to noise (first term in $D$) is negligible:
\begin{equation}
    J \simeq \frac{1}{2\pi} \int d\omega\,\mathcal{T}(\omega),
\ D \simeq \frac{1}{2\pi} \int d\omega\, \mathcal{T}(\omega)
\left[ 1-\mathcal{T}(\omega) \right] \label{eq:flatten_JD}
\end{equation}
and the Fano factor, $F=1- \int d\omega\,\mathcal{T}^2(\omega)/\int d\omega\,\mathcal{T}(\omega)$. Thus, in this regime, a low Fano factor requires a transmission profile that is close to one in the transport window and close to zero outside the window. Simply narrowing the window is not sufficient, because it also reduces the mean current. The useful regime is instead a broad, flat, high-transmission plateau with minimal sharp features. Rectangular or boxcar-like transmissions are known to be favorable for fluctuation suppression in coherent two-terminal conductors~\cite{ehrlich2021broadband}. As a systematic extension of this principle, we now explain a strategy for current noise suppression by optimizing over the reservoir bandwidth and harnessing advantages over the Markovian limits of the thermodynamic and kinetic precision coefficients.

Since the residual Markovian reservoirs couple only to the two auxiliary boundary modes of the tight-binding chain of the extended chain of length $N+2$ in the pseudomode formalism [Eq.~\eqref{eq:pseudo_ham}], the Fisher--Lee relation gives [Appendix~\ref{ap:inter}]
\begin{equation}
\mathcal{T}(\omega)=\eta^2\left|G^r_{1,N+2}(\omega)\right|^2=\eta^2\left|\sum_j \frac{A_j}{\omega-\widetilde{\omega}_j}\right|^2 \label{eq:pole_expansion}
\end{equation}
with the poles $
\widetilde{\omega}_j = \Omega_j - i\gamma_j/2$, where $\Omega_j$ is the resonance frequency of mode $j$ and $\gamma_j$ is its dissipation. $A_j$ is the residue of pole $j$, which sets how strongly that mode contributes to transport. The poles of end-to-end propagator $G^r(\omega)_{1,N+2}=\langle 1|G^r(\omega)|N+2\rangle$ of the retarded Green's function $G^r(\omega)= 
[\omega I-H_{\rm ext}+\frac{i}{2}\eta( |1\rangle\langle 1|+ |N+2\rangle\langle N+2|)]^{-1}$, therefore determine the full frequency dependence of the transmission. 

The transmission function in Eq.~\eqref{eq:pole_expansion} naturally splits into two types of contributions. The direct terms, leading to individual squared Lorentzians, are strictly positive and symmetric, and thus they can not produce non-monotonic behavior by themselves. The minimum in the Fano lineshape instead originates from the cross-terms $A_jA_k^\ast$, which encode quantum interference between pairs of transport modes \cite{fano1, clerk2001, fanoresonance}. These interference contributions survive energy integration only when the pole broadenings $\gamma_j$ are comparable to the inter-pole frequency splittings $\Omega_j-\Omega_k$ [Appendix~\ref{ap:inter}].  In this case, constructive interference flattens the transmission window, thereby suppressing the noise contribution $D$ [Eq.~\eqref{eq:flatten_JD}] without strongly suppressing the current, and therefore generates the Fano-factor minimum visualized in Figs.~\ref{Fig2}(a1--a3, b). The results shown are calculated using exact nonequilibrium Green's function (NEGF) techniques [Appendix~\ref{ap:negf}]. These observations suggest a systematic two-step strategy for minimizing the current Fano factor. 

\textit{Step 1: set internal mode spacing and damping.---}A chain of $N$ dots with uniform hopping $g$ has a tight-binding spectrum $E_k = 2g\, \cos(k\pi/(N+1))$, spanning a bandwidth of $4g$ \cite{kittel2004introduction, ashcroft1976solid}. For the boundary dissipation $\Gamma$ to induce transparent transport, it must match the internal hopping $\Gamma = 2g.$
This is the known Markovian impedance-matching condition~\cite{markovian}, as discussed in Appendix \ref{ap:crit}. For a tight-binding chain, this condition guarantees $\mathcal{T}(0) = 1$ regardless of the reservoir bandwidth $\eta$, confirming it is a necessary prerequisite. An exception to the optimality of this condition is discussed in  Appendix~\ref{ap:transmission}.

\textit{Step 2: match bath and chain bandwidths.---}For Lorentzian baths, the boundary dissipation becomes frequency dependent and decreases away from resonance over the bandwidth scale $\eta$. The interference cross-terms in Eq.~\eqref{eq:pole_expansion} remain effective after the Landauer energy integration only if the reservoir bandwidth covers the relevant internal spectrum $E_k$ of the chain, so that the off-center transport resonances remain sufficiently damped. Combining this bandwidth-coverage condition with the reflectionless boundary condition in the pseudomode representation gives [Appendix~\ref{ap:Nchainscale}]
\begin{equation}
\eta_{\mathrm{opt}}=\frac{8g^2}{\Gamma}.
\label{eq:theequation}
\end{equation}
This is the non-Markovian memory matching condition for Lorentzian baths. It specifies that the optimal reservoir correlation time $(\sim 1/\eta)$ must be finite. A Markovian bath ($\eta\rightarrow\infty$) fails because it cannot provide the frequency selectivity needed to sustain coherent multi-mode interference, whereas a very narrow bath ($\eta\rightarrow 0$) suppresses transport by creating a memory-induced bottleneck [Appendix~\ref{app:memory-induced}]. Evaluating Eq.~\eqref{eq:theequation} at the Markovian condition $\Gamma=2g$ yields $\eta_{\rm opt}=4g$, equal to the tight-binding chain bandwidth~\cite{kittel2004introduction, ashcroft1976solid}, thus confirming that at the impedance-matched point the bath bandwidth precisely covers the internal spectrum.

We name the two-step strategy above as \emph{dual impedance-matching}. It systematically synchronizes three characteristic scales of the problem: the internal mode splitting ($\sim g$), the boundary dissipation rate ($\sim \Gamma$), and the bath bandwidth ($\sim\eta$). Our calculations show excellent agreement with the predictions of the above mechanism, as shown in Fig.~\ref{Fig2}(b).

For the double quantum dot in the high-bias limit, the same mechanism gives a finite-bandwidth maximum of the mean current, $\eta_{\mathrm{opt}}^{J}=2g\left(2g+\sqrt{4g^2+\Gamma^2}\right)/\Gamma$, showing that the current is not generally maximized in the Markovian wide-band limit [Appendix~\ref{ap:dqd}]. This current optimum does not exactly coincide with the Fano-factor minimum $\eta_{\rm opt}$, because minimizing $F=D/J$ requires suppressing noise while maintaining transmission. Nevertheless, both optima lie in the same finite-memory window, and the mechanism persists for longer chains, with only finite-size and parity-dependent shifts of the detailed Fano lineshape [Appendix~\ref{ap:ndot}].

The globally optimal internal hopping in the finite-memory regime differs slightly from the Markovian value implied by $\Gamma=2g$. This shift has a simple origin. A structured reservoir provides frequency-dependent damping and energy renormalization. If $g$ is kept at its Markovian value, the finite-chain modes away from the band center can be pushed into spectral regions where the bath provides insufficient damping, producing sharp peaks and valleys in $\mathcal{T}(\omega)$ that increase the Fano factor [Fig.~\ref{Fig2}(c)]. A slightly smaller $g$ pulls these modes back into the effective reservoir bandwidth, compensating the renormalization effects, thereby producing a flatter transmission window and lower noise. The matching condition also provides a practical route to the optimum without requiring a full two-dimensional search in the $(g,\eta)$ plane. As shown in Fig.~\ref{Fig2}(d), the two-step procedure, first fixing the near-Markovian impedance-matched coupling and then optimizing the reservoir bandwidth, closely tracks the global minima obtained by unconstrained optimization. The small deviation of the global optimum reflects the additional balance between level repulsion, finite-bandwidth noise filtering, and the mean current.

\begin{figure}
    \centering
    \includegraphics[trim=.2cm 0.2cm 0.2cm 0.cm, clip,width=\linewidth]{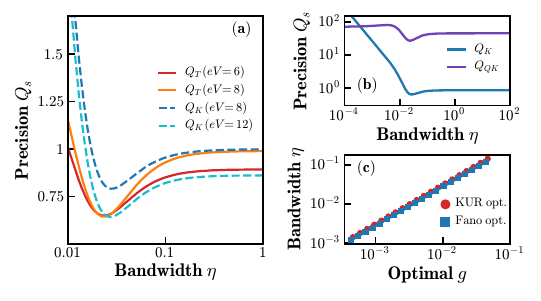}
    \parbox{0.49\textwidth}{
    \caption{\justifying
    Thermodynamic and kinetic precision enhancement using bath memory.
    (a) Simultaneous reduction of TUR and KUR coefficients for different values of bias. 
    (b) Semiclassical KUR and quantum kinetic coefficients as functions of reservoir bandwidth for $eV/T=8$. 
    (c) Global minima of semiclassical KUR and Fano factor in the $(g,\eta)$ plane for $\Gamma \in [0.001,0.1]$.
    Parameters: $\Gamma_\alpha=0.01$, $T=1$, $\varepsilon_{L_\alpha}=\varepsilon=1$, $N=2$.
    \label{fig:simultaneo}}}
\end{figure}

\textit{\textbf{Memory-assisted thermokinetic precision.---}}We now connect the finite-memory noise suppression discussed above to thermodynamic and kinetic precision bounds for nonequilibrium transport fluctuations, which are generally expressed using inequalities of the form 
\begin{equation}
\mathcal{Q}_s = \frac{\mathcal{C}_s D}{J^2}\geq 1.
\label{eq:gen_defn_ur}
\end{equation}
One of the most well-studied bounds of the above form is the thermodynamic uncertainty relation (TUR) \cite{Turra, markovian} that bounds the coefficient $\mathcal{Q}_T\geq 1$ with $\mathcal{C}_T=\sigma/2$. It features the entropy production rate which can be expressed as $\sigma= eV\beta J$  for reservoirs with equal temperatures. Another important bound is the kinetic uncertainty relation (KUR), \cite{KUR} $\mathcal{Q}_K\geq 1$, that uses the dynamical activity $K$ through $\mathcal{C}_K=K$. The dynamical activity measures the overall rate of microscopic transport events between the system and the reservoirs.
For the double-dot system, we use the semiclassical activity~\cite{markovian}, $K = K_{\rm bound} + K_{\rm int}$ [Appendix~\ref{ap:K}], where $K_{\rm bound}$ accounts for particle exchanges between the auxiliary lead modes and the boundary dots, and $K_{\rm int}$ accounts for coherent internal hopping in the double dot. Since $K$ is sensitive to quantum coherence, $Q_K$ can violate the classical bound of unity for transport scenarios where coherence can reduce noise $D$ without increasing $K$ proportionately. We also study the quantum kinetic uncertainty relation (QKUR)~\cite{blasi2026quantumkineticuncertaintyrelations},  $\mathcal{Q}_{QK}\geq 1$ with $\mathcal{C}_{QK}=K_Q$,
where $K_Q$ is constructed directly from scattering quantities  [Appendix~\ref{ap:KQ}], unlike the semiclassical activity $K$ above, which acts as a measure of transitions. Thus, $\mathcal{Q}_{QK}$ remains strictly bounded even when quantum transport overcomes the classical precision bound.

In the Markovian limit $(\eta\to\infty)$, our calculations  reproduce the double-dot Markovian \textit{benchmark} of Ref.~\cite{markovian}, as shown in Appendix~\ref{appendixc}. In this limit, $\mathcal{Q}_K$ is reduced at high voltage bias (where $J$ dominates over $K$), whereas $\mathcal{Q}_T$ is reduced at lower bias (where $\sigma$ is smaller). The result, as also reported in the Markovian limit results of Ref.~\cite{markovian}, is that only a weak simultaneous thermodynamic and kinetic precision advantage is possible near $eV/T\sim 8$ [Fig.~\ref{fig:simultaneo}(a)]. We show that the finite bath memory $\eta$ can act as an additional control parameter for transport and provide stronger precision advantage over the classical thermodynamic and kinetic bounds simultaneously.

Figure~\ref{fig:simultaneo}(a) shows this joint reduction of both precision coefficients near the dual impedance-matching scale across a range of bias voltages. Figure~\ref{fig:simultaneo}(b) compares $\mathcal{Q}_K$ with $\mathcal{Q}_{QK}$ as functions of $\eta$. Both display a pronounced finite-memory minimum at very similar bath bandwidths, indicating that the improvement is not merely an artifact of the semiclassical activity estimate, but a consequence of the quantum interference mechanism identified in our discussion of the impedance-matching conditions introduced above. Note that although the semiclassical KUR coefficient accurately reproduces the relevant finite memory improvement, it diverges in the limit ($\eta \rightarrow 0$) reflecting low current bottleneck in that regime. Figure~\ref{fig:simultaneo}(c) shows that the global kinetic precision coefficient and Fano factor minima track each other  across the scanned values of hopping strength $g$, thus linking the kinetic-precision improvement to the same noise-suppression mechanism identified from the Fano-factor lineshape. These observations validate the utility of the semiclassical activity metric.

\textit{\textbf{Generalization to arbitrary baths.---}}The scaling relation in Eq.~\eqref{eq:theequation} is specific to Lorentzian baths. However, we now show that the underlying physical mechanism of noise suppression through bath bandwidth matching is a structural feature of quantum transport, and persists for other bath structures. A key quantity of interest is the dissipation that the Lorentzian density $\mathcal{J}(\omega, \eta_{\rm opt})$  [Eq.~\eqref{eq:lorentzian_density_results}] with the optimum bandwidth $\eta_{\rm opt}$ [Eq.~\eqref{eq:theequation}] provides at the modes near the boundary of the transmission band.  For an arbitrary structured density $\mathcal{J}_X(\omega)$, a scaling relation similar to Eq.~\eqref{eq:theequation} can be derived through the following procedure.

We first identify a representative bandwidth (or cutoff frequency) $\omega^{\text{base}}_c$ that equates the spectral density of the arbitrary bath $\mathcal{J}_X(\omega_r, \omega_c^{\text{base}})$ to the optimal target dissipation in the Lorentzian case,  $\mathcal{J}(\omega_r, \eta_{\rm opt})$. The frequency $\omega_r = 2g \cos(\pi/(N+1))$ is the boundary mode for a tight-binding chain. For baths with Gaussian ($\mathcal{J}_G(\omega)=\Gamma e^{-(\omega/\omega_c)^2}$) and Newns ($\mathcal{J}_N(\omega)=\Gamma \sqrt{1-(\omega/\omega_c)^2}$) spectral densities, a simple algebraic inversion yields the representative bandwidth $\omega^{\text{base}}_{c}$. The details of the calculations can be found in Appendix~\ref{ap:other}. 

Any structured bath imprints a complex self-energy $\Sigma_X(\omega)=\Delta_X-\frac{i}{2}\mathcal{J}_X(\omega)$, where the imaginary part sets the mode dissipation and the real part, the Lamb shift contribution $\Delta_X(\omega) = \frac{1}{2\pi} \mathcal{P} \int d\omega'\mathcal{J}_X(\omega')/(\omega - \omega') $, displaces the resonant poles from the optimum. Thus, matching the dissipation alone is not sufficient. At $\omega^{\text{base}}_{c}$, the Lamb shift displaces the boundary mode to $\omega_{r,X} =\omega_r + \Delta_X(\omega_r, \omega_c^{\text{base}})$, whereas the Lorentzian reference places it at $\omega_{r,L} =\omega_r + \Delta_L(\omega_r, \eta_{\text{opt}})$. To restore the matching condition of the interference cross-terms, we rescale the bandwidth by the ratio of the two renormalized frequencies to estimate the optimum bandwidth for the spectral density $\mathcal{J}_X(\omega)$:
\begin{equation}
    \omega_c^{\text{opt}} = \omega_c^{\text{base}} \frac{\omega_{r,L}}{\omega_{r,X}}.
    \label{eq:scaling_ansatz}
\end{equation}
The respective Lamb shifts are calculated in Appendix~\ref{ap:other}. As shown in Fig.~\ref{baths}, this procedure accurately predicts the local noise minima for both the smooth Gaussian spectral density and the hard-cutoff Newns spectral density.

Fig.~\ref{baths}(b) shows agreement of the results even when thermal fluctuations become dominant in the system. As expected, when the bias becomes too weak to support constructive interference between the transmission modes, the pronounced nonmonotonic Fano lineshape is lost. However, even in this regime, we see that finite memory can induce finite albeit negligible transport noise suppression over its  Markovian wide-band limit.

\begin{figure}[t]
    \centering
    \includegraphics[width=0.48\textwidth]{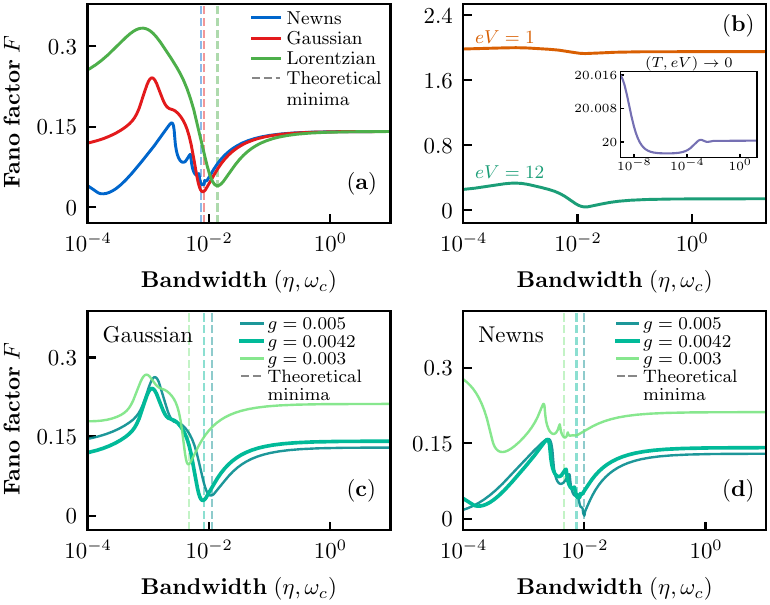}
    \parbox{0.48\textwidth}{
    \caption{\justifying Fano factor $F$ vs effective bandwidth $\eta$ (Lorentzian) and $\omega_c$ (Gaussian and Newns). Vertical dashed lines indicates the theoretically predicted minima [Eq.~\eqref{eq:scaling_ansatz}]. (a) $F$ for different bath spectra for $g=0.005$. (b) $F$ for different values of bias $eV$; the inset shows the linear response behavior for $T=10^{-3}$, $eV=10^{-4}$. The Fano factor $F$ with respect to bandwidth is shown for different $g$'s for a Gaussian (c) and a Newns (d) bath spectral density. Parameters: $N=10$, $\Gamma_\alpha=0.01$, $T=1$, $eV/T=12$, and  $\varepsilon_{L_\alpha}=\varepsilon=1$, unless otherwise stated.
    \label{baths}}}
\end{figure}

\textit{\textbf{Discussion and outlook.---}}We demonstrate that, for fixed system parameters, a structured bath with a finite memory timescale suppresses the current Fano factor below its Markovian value via a dual impedance-matching mechanism. We establish that, by exploiting the bath structure as a control parameter for precision enhancement, the current Fano factor can be minimized concurrently with both the thermodynamic and kinetic uncertainty coefficients. The mechanism is captured by a closed-form analytical scaling law [Eq.~\eqref{eq:theequation}] and can predict the optimal condition accurately for Lorentzian, Gaussian, and Newns spectral densities through a Lamb-shift renormalization.

This finding motivates a concrete practical application of the proposed matching conditions. A systematic sweep of the system parameters, such as the interdot tunneling, dot–lead coupling strength, or the reservoir bandwidth reveals the parameters for optimal transport precision  through the minimum of the Fano factor [Fig.~\ref{fig:simultaneo}(c)] without requiring reconstruction of the dynamical activity, a task that is typically more demanding than current–noise measurements~\cite{KUR, markovian, blasi2026quantumkineticuncertaintyrelations}. Moreover, our explicit scaling relations for Lorentzian environments, together with their extensions to more general structured baths, provide a practical blueprint for engineering optimal transport regimes in realistic quantum platforms. Our results can be tested in existing platforms that support coherent transport, including semiconductor quantum-dot arrays~\cite{vanderWiel_2002, hensgens2017quantum, volk2019loading, mills2019shuttling} and ultracold fermionic gases~\cite{Brantut_2012, brantut2013thermoelectric, krinner2015observation, PhysRevX.8.011053}.

\section*{Acknowledgements} \label{sec:acknowledgements}
 This work was supported by the QuantERA II program (Mf-QDS) and QuantERA III program (AQuSeND) that have received funding from the European Union’s Horizon 2020 research and innovation program under Grant Agreement No.101017733 and from the Agencia Estatal de Investigación, with project codes PCI2022-132915, PCI2024-153474 and by QNAVIUM Project SCPP2400C011413XV0 funded by MICIU/AEI/10.13039/501100011033 and by the European Union NextGenerationEU/PRTR. And by the European Union (Quantum Flagship project ASPECTS, Grant Agreement No. 101080167).

\nocite{Turra}
\bibliographystyle{apsrev4-2} 
\bibliography{bibliography}

\onecolumngrid

\appendix
\section*{\textbf{Supplementary material}}

\section{Metrics and definitions}

\subsection{Lorentzian spectral density and pseudomodes}
\label{ap:pseudo}
Lorentzian spectral densities provide a minimal structured-bath model with a finite correlation time and are often used as building blocks for more complex environments~\cite{Mark,PhysRevA.80.012104}. They therefore constitute a natural first step beyond the Markovian wide-band limit. We use
\begin{equation}
    \mathcal{J}(\omega) = \frac{ \eta^2 \Gamma}{4{(\omega - \epsilon_{L})^2} + \eta^2},
\end{equation}
where $\omega$ is the frequency, $\eta$ is the spectral width, and $\Gamma$ is the on-resonance value. For the noninteracting problem studied here, this spectral density can be implemented directly within NEGF (Appendix~\ref{ap:negf}). Equivalently, the Lorentzian reservoir can be represented through a Markovian extension using the pseudomode formalism~\cite{Mark,PhysRevA.80.012104}: the structured reservoir is replaced by an explicit auxiliary lead mode damped by a residual Markovian bath at rate $\eta$. The extended Hamiltonian is
\begin{equation}
    H_{\text{ext}}=H_{\text{sys}}+\sum_{\alpha=L}^R \varepsilon_{\alpha}c_{\alpha}^\dagger c_{\alpha}+ \left(\lambda_L c_{L}c^\dagger_{1} +\lambda_R c_{R}c^\dagger_{N} + \text{H.c.} \right)\label{eq:pseudo_ham}
\end{equation}

\begin{figure}[b]
    \centering
    \includegraphics[width=.7\linewidth]{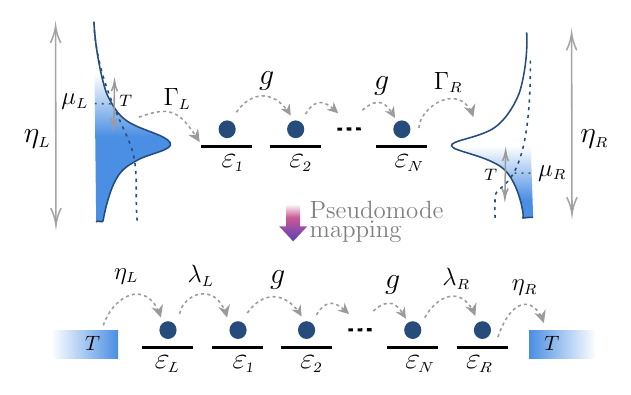}
    \parbox{\textwidth}{
    \caption{\justifying Diagram of the $N$-site chain system.  The internal coupling strength between dots is given by $g$, $\lambda_i$ is the coupling between the leads and the boundary dots, and $\eta_i$ is the dissipation rate coupling the leads to the residual macroscopic baths which is identifiable with memory \cite{PhysRevA.80.012104}.
    \label{system}}}
\end{figure}

The system is represented in the diagram shown in Fig.~\ref{system}.

The coupling between the boundary dot and the auxiliary mode is chosen so that the effective spectral density seen by the dot has the same on-resonance amplitude $\Gamma$. Each end of the chain sees an effective Lorentzian spectral density given by \cite{Mark, PhysRevA.80.012104}:

\begin{equation}
\label{espectro}
J_{\text{eff}_\alpha}(\omega) = 2\lambda_\alpha^2\frac{ (\frac{\eta_\alpha}{2})}{(\omega - \epsilon_{L_\alpha})^2 + (\frac{\eta_\alpha}{2})^2},
\end{equation}

To keep $J_{\text{eff}_\alpha}$ constant and equal to $\Gamma_\alpha$ in resonance, we must fix resonance conditions $\varepsilon_{D_\alpha}=\varepsilon_{L_\alpha}$ and solve for $\lambda_\alpha$ enforcing $J_{\text{eff}_\alpha}= \Gamma_\alpha$, this yields:

\begin{equation}
\label{eq:rc}
\lambda_\alpha=\sqrt{\left(\frac{\eta_\alpha^2}{4} + (\varepsilon_D{_\alpha}-\varepsilon_{L_\alpha})^2\right)\frac{\Gamma_\alpha}{\eta_\alpha}}.
\end{equation}

At resonance this gives $2\lambda_\alpha=\sqrt{\eta_\alpha\Gamma_\alpha}$. For clarity, we focus on symmetric reservoirs and write $\Gamma_L=\Gamma_R=\Gamma$ and $\eta_L=\eta_R=\eta$ unless stated otherwise.

This setup allows us to tune the environment from the strong-memory regime (small $\eta$) to the Markovian limit without additional approximations~\cite{Mark}. For $\eta\to\infty$, the bath correlation time $\eta^{-1}$ becomes much shorter than the intrinsic system timescales, and the standard Markovian results are recovered.

\subsection{Memory-induced bottleneck measure}
\label{app:memory-induced}
Rigorous quantification of non-Markovianity typically requires evaluating the information backflow via time-evolution metrics, such as the Breuer-Laine-Piilo (BLP) measure \cite{Breuer09}. However, evaluating such dynamical map properties is computationally prohibitive and physically disconnected from the nonequilibrium steady-state transport studied here. Instead, within the mesoscopic reservoir formalism, memory effects manifest structurally as the deviation of the auxiliary lead modes from thermal equilibrium, thus allowing us to measure memory impact on the system based on the ratio:

\begin{equation*}
\mathcal{M}_{r}=\frac{N_L+f_R}{N_R+f_L}
\end{equation*}

As defined, $\mathcal{M}_{r}$ reaches $1$ in the Markovian limit ($N_{\{L,R\}} \to f_{\{L,R\}}$) and decreases as the bath correlation times increase, acting as an indicator for the memory-induced bottleneck in our setup. Using Eq.~\ref{ec:ocup}, this quantity admits a simple form for symmetric bath memory and symmetrically applied bias, $\mu_{L,R}=\varepsilon_{L,R} \pm eV/2$:

$$
\mathcal{M}_{r}=\frac{\eta - J}{\eta + J}.
$$

This quantity may also be given in terms of the correlation time $\tau_{\text{bath}}$:
$$
\frac{1-J\tau_{\text{bath}} }{1+J\tau_{\text{bath}} }
$$
Thus, within the resonant pseudomode description, $\mathcal{M}_r$ provides a direct measure of the memory-induced bottleneck. Although it depends on the applied bias before high-bias saturation, we show below that it remains stable against changes in $\Gamma$ for a fixed chain length. It therefore provides a useful diagnostic for the Fano-factor minima.

\subsection{Semiclassical Dynamical Activity Definition}
\label{ap:K}

In this article, we employed a semiclassical model for dynamical activity, as described in \cite{markovian}. This involves approximating the impact of Rabi oscillations between quantum dots and summing this quantity to the conventional boundary dynamical activity. This quantity is limited to the DQD as longer chains are computationally intractable due to the scaling of multi-site internal coherence.

A naive extension of the boundary activity to the extended system would lead to thermodynamic inconsistencies. One might write, by analogy with the Markovian definition,

$$
K_{\text{bound}}= \Gamma(f_L(1-N_D) + N_D(1-f_L))
$$

Within the semiclassical construction of Ref.~\cite{markovian}, this term corresponds to coherent hopping between the boundary dot and the auxiliary pseudomode, constrained by Eq.~\ref{eq:rc}. The corresponding effective transition rate is

$$
\mathcal{W}_{\text{L,D}}= \frac{4\lambda^{2}}{\eta}=\Gamma
$$

where the last equality follows from Eq.~\ref{eq:rc} at resonance. Away from the Markovian limit, correlations between the auxiliary modes and boundary dots are generally non-negligible. Using Wick's theorem and extending the definitions of Ref.~\cite{markovian} to the enlarged system, the boundary contribution becomes

\begin{equation}
\label{actividadnomarko}
K_{\text{bound}} = \sum_{i=\text{1,2}} \Gamma \ (N_{L_i}(1-N_{D_i})+N_{D_i}(1-N_{L_i}) + 2|\rho_{L_iD_i}|^2)
\end{equation}

In the $\eta\to\infty$ limit the auxiliary-mode coherences vanish, recovering the usual Markovian expression,

$$
K_{\text{bound}} = \sum_{i=\text{1,2}} \Gamma \ (f_i(1-N_{D_i})+N_{D_i}(1-f_i))
$$

We must also account for coherence between dots, this is penalized in the DQD case as \cite{markovian}:
$$
K_{\text{int}}= \sum_{i=1}^2 \frac{4g^2}{\Gamma_1 + \Gamma_2} \langle N_{D_i}\rangle - \frac{8g^2}{\Gamma_1 + \Gamma_2} \langle N_{D_1}N_{D_2} \rangle
$$
For consistency, we will use Eq.~\ref{actividadnomarko} to penalize coherences induced by memory and establish comparisons with more modern definitions, such as the Quantum Kinetic Uncertainty Relation (QKUR) in the DQD setup \cite{blasi2026quantumkineticuncertaintyrelations} to check for possible artifacts emerging from the semiclassical approximation. After the comparisons between metrics are studied, we can then directly measure the QKUR in longer chains to assess advantage.

\subsection{Quantum dynamical activity definition}
\label{ap:KQ}

We have also considered the quantum kinetic uncertainty relation:
\begin{equation}
\mathcal{Q}_{QK}
=
\frac{D K_Q}{J^2}\geq 1,
\label{eq:QQK_definition_results}
\end{equation}
where~\cite{blasi2026quantumkineticuncertaintyrelations}
\begin{equation}
K_Q
=
\frac{\mathcal{A}_{\rm cross}^2}
{\mathcal{A}_{\rm cross}-\mathcal{A}_{\rm sh}}.
\end{equation}
The two quantities entering $K_Q$ are
\begin{equation}
\mathcal{A}_{\rm cross}
=
\frac{1}{2\pi}
\int_{-\infty}^{\infty}
d\omega\,
\mathcal{T}(\omega)
\left[
f_L(1-f_R)+f_R(1-f_L)
\right],
\end{equation}
and
\begin{equation}
\mathcal{A}_{\rm sh}
=
\frac{1}{2\pi}
\int_{-\infty}^{\infty}
d\omega\,
\left[
1-\mathcal{T}(\omega)
\right]
\left[
f_L-f_R
\right]^2.
\end{equation}

This quantity is defined purely from scattering theory principles, and the bound it poses remains strict for every quantum system \cite{blasi2026quantumkineticuncertaintyrelations}. Thus, QKUR can be used as a check for possible artifacts in the KUR.

\section{The critical damping condition}
\label{ap:crit}
In this section we consider a symmetrically coupled DQD with $\varepsilon_{D_1}=\varepsilon_{D_2}=\varepsilon_\alpha$ for simplicity.
The transmission probability derived from the Green's function is given by \cite{datta2005quantum}:

\begin{equation}
\label{tfunction2}
    \mathcal{T}(\omega) = \frac{4g^2 (\text{Im}[\Sigma(\omega)])^2}{|(\omega - \varepsilon_\alpha - \Sigma(\omega))^2 - g^2|^2}
\end{equation}

To find the resonance condition that maximizes the transmission we define $x = \omega -\varepsilon_\alpha - \text{Re}[\Sigma(\omega)]$ and $y = -\text{Im}[\Sigma(\omega)]$. Because the self-energy must be dissipative, $y \ge 0$. Substituting these variables into the maximized transmission function yields:
\begin{equation}
    \frac{4g^2 y^2}{(x^2 - y^2 - g^2)^2 + 4x^2y^2} = 1.
\end{equation}

The expression simplifies to $(x^2 + y^2 - g^2)^2 = 0 \implies x^2 + y^2 = g^2$. Reverting to our original complex variables, we note that $x^2 + y^2 = |\omega -\varepsilon_\alpha- \Sigma(\omega)|^2$. This provides the exact condition for perfect transmission in the effective single-particle problem:
\begin{equation}
\label{eq:geometric}
    |\omega -\varepsilon_\alpha - \Sigma(\omega)|^2 = g^2
\end{equation}

We can evaluate Eq.~\ref{eq:geometric} at the central resonance frequency $\tilde\omega_0 = \varepsilon_\alpha+ \text{Re}[\Sigma(\tilde\omega_0)]$. At this point, the condition reduces to:
\begin{equation}
\label{condicion0}
-\text{Im}[\Sigma(\tilde\omega_0)] = g
\end{equation}

Since the internal splitting for this $N=2$ case is $\Delta\Omega=2g$ and the effective collective damping is $\bar{\gamma}=-2\text{Im}[\Sigma]$, we recover $\bar{\gamma}=|\Delta\Omega|$. For Eq.~\ref{condicion0} to be achievable, the spectral density at resonance must satisfy $J(\tilde\omega_0)=-2\mathrm{Im}[\Sigma(\tilde\omega_0)]=2g$. There is therefore a dissipation threshold below which perfect transmission cannot be reached. For a Markovian bath with dissipation strength $\Gamma$, this gives $\Gamma=2g$, the critical-damping condition reported in previous Markovian analyses~\cite{markovian}. Because this condition is imposed pointwise at resonance, it does not necessarily optimize the transmission over a finite window in a non-Markovian system, but it provides a useful estimate for the parameters discussed in Appendix~\ref{ap:transmission}.

The resonance condition is controlled by the matching between the internal coupling $g$ and the complex detuning induced by the bath. If the spectral density entering $\Sigma(\omega)$ is too small, Eq.~\ref{eq:geometric} may have no real solution within the transport window.

\section{Nonequilibrium Green's functions (NEGF) techniques}
\label{ap:negf}
The NESS transport properties are calculated using the nonequilibrium Green's function (NEGF) formalism applied to the extended system. For the noninteracting model considered here, this formulation is exact. For Lorentzian baths, we implement the pseudomode representation rather than the continuous Lorentzian self-energy because it is computationally simple and directly comparable with the analytical construction.

The retarded Green's function is given by \cite{Landauer, markovian}:

$$
\mathcal{G}^r(\omega)=\frac{1}{\omega -H_{ext} - \Sigma_{res}}
$$

Here $H_{\rm ext}$ is the extended Hamiltonian, including the auxiliary lead modes. The residual Markovian baths contribute the self-energy $\Sigma_{\rm res}=-i\eta/2$ on the auxiliary modes.

The transmission function is defined by the Fisher-Lee formula \cite{Landauer}:
\begin{equation}
\mathcal{T}(\omega)=\mathrm{Tr}[\Gamma_L \mathcal{G}^r(\omega)\Gamma_R\mathcal{G}^a(\omega)]
\end{equation}
Here, $\mathcal{G}^{r,a}$ are the Green's functions of the extended system, and $\Gamma_\alpha=i(\Sigma_\alpha^r- \Sigma_\alpha^a)$.

The current $J$ is obtained from the Landauer--Büttiker formalism \cite{Landauer}:
$$
J= \frac{1}{2\pi} \int_{-\infty}^\infty \mathcal{T}(\omega)[f_L(\omega)- f_R (\omega)] d\omega
$$
The zero-frequency noise $D$ is given by
\begin{equation}
    D 
= 
\frac{1}{2\pi}
\int_{-\infty}^{\infty}d\omega\,
\Big\{
\mathcal{T}(\omega)
\left[
f_L(\omega)(1-f_L(\omega))+f_R(\omega)(1-f_R(\omega))
\right]
+\mathcal{T}(\omega)
\left[
1-\mathcal{T}(\omega)
\right]
\left[
f_L(\omega)-f_R(\omega)
\right]^2
\Big\},
\end{equation} 
where $ f_\alpha (\omega)= [e^{\frac{\omega- \mu_\alpha}{T}}+1]^{-1}$ are the Fermi-Dirac distributions of the residual baths.

To evaluate the KUR in the DQD case we also require the effective dynamical activity $K$ at the interface between the auxiliary leads and the system. This quantity depends on the stationary population of the lead modes, $N_L$, which generally deviates from the thermal distribution due to memory effects. The quantity $N_L$ can be derived exactly from the steady-state current-conservation law at the lead sites. The net current entering the lead mode from the residual bath, $I_{bath \to L} = \eta (f_L(\epsilon_L) - N_L)$, must equal the current flowing into the system, $J$. This yields the relation used, for the sake of simplicity and stability, in our numerical implementation:

\begin{equation}
\label{ec:ocup}
N_{L,R} = f_{L,R}(\epsilon_{L,R}) \mp \frac{J}{\eta}.
\end{equation}

Here, $f_{L,R}(\epsilon_{L,R})$ are the Fermi-Dirac distributions evaluated at the auxiliary-mode energies; the upper/lower sign corresponds to the left/right lead.

We must also calculate $\langle N_1N_2\rangle$. Using Wick's theorem~\cite{markovian},
\begin{equation}
\langle N_1N_2\rangle= \langle N_1\rangle \langle N_2\rangle - |\rho_{12}|^2
\end{equation}

where $\rho_{12}$ is the corresponding one-body density-matrix element.

\section{Origin of the Fano minimum}
\label{ap:inter}
\subsection{The double quantum dot case}
\label{ap:dqd}

To understand why sweeping the Lorentzian width $\eta$ reveals optimality, we must analyze the analytical structure of the transmission function and its survival upon integration.

In the basis of the auxiliary leads and the quantum dots, $\{ |L_1\rangle, |D_1\rangle, |D_2\rangle, |L_2\rangle \}$, the effective single-particle Hamiltonian of the open system is given by:

\begin{equation}
H_{\text{eff}} =\begin{pmatrix}\epsilon_{L} - i\frac{\eta}{2} & \lambda & 0 & 0 \\  \lambda & \epsilon_{D1} & g & 0 \\ 0 & g & \epsilon_{D2} & \lambda \\ 0 & 0 & \lambda & \epsilon_{R} - i\frac{\eta}{2}\end{pmatrix}
\end{equation}

Because the left and right residual baths only interact with the auxiliary leads, the matrices $\Gamma_{L,R}$ are diagonal with only a single non-zero element ($\eta$) at positions $(1,1)$ and $(4,4)$~\cite{datta2005quantum}, respectively. Multiplying these matrices inside the trace yields:
\begin{equation}
\label{ec:trans}
\mathcal{T}(\omega) = \eta^2 |\mathcal{G}^r_{14}(\omega)|^2
\end{equation}

The transport behavior is thus encoded in a single off-diagonal element of the retarded Green's function, $\mathcal{G}^r = (\omega I - H_{\text{eff}})^{-1}$. By Cramer's rule, the inversion of this matrix gives $\mathcal{G}^r_{14}(\omega)$ as the $(4,1)$ cofactor divided by the determinant. Because the system is a 1D chain, the submatrix for this cofactor is lower triangular. Its determinant is simply the product of the off-diagonal coupling constants $(-\lambda)(-g)(-\lambda) = -g\lambda^2$. Therefore, there is no $\omega$ dependence in the numerator. The dispersion of the transmission $\mathcal{T}(\omega) = \eta^2 |\mathcal{G}^r_{14}(\omega)|^2$ is fully dictated by the characteristic polynomial in the denominator, $\det(\omega I - H_{\text{eff}})$. By evaluating the squared modulus of this determinant and imposing $\lambda^2 = \frac{\eta\Gamma}{4}$, the transmission takes the form:

\begin{equation}
\begin{split}
\mathcal{T}^{(N=2)}(\omega) = 
\frac{\eta^4 \Gamma^2 g^2}{16 \ D(\omega)} 
\end{split}
\end{equation}
where $D(\omega)$ is
\begin{equation*}
\begin{split}
D(\omega) = 
&\left[
\omega^4 
- \left(g^2 + \frac{\eta^2}{4} + \frac{\eta\Gamma}{2}\right)\omega^2 
+ \frac{\eta^2}{4}\left(g^2 + \frac{\Gamma^2}{4}\right)
\right]^2 \\
&+ \eta^2\omega^2 
\left[
\omega^2 
- \left({g^2} + \frac{\eta\Gamma}{4}\right)
\right]^2
\end{split}
\end{equation*}

This explicitly shows how the memory reshapes the transmission. Notably, evaluating this expression at $\omega=0$ factors out the memory dependence, yielding:

\begin{equation}
\mathcal{T}^{(N=2)}(0) = \frac{16 \Gamma^2 g^2}{\left( 4g^2 + \Gamma^2 \right)^2}
\end{equation}

This limit confirms that $\mathcal{T}(0) \to 1$ when the Markovian impedance-matching condition ($\Gamma = 2g$) is satisfied, regardless of the memory timescale. Extracting exact results requires computing the full transport integrals. In the high-bias regime ($eV \gg T$), the current simplifies to $J_{\text{max}} = \frac{1}{2\pi} \int_{-\infty}^{\infty} \mathcal{T}(\omega) d\omega$. To avoid numerical integration of the 8th-degree rational function, we can apply the James-Nichols-Phillips integral formulas to its 4th-degree complex denominator \cite{james1947theory}.

To apply this formulation, we map the frequency domain to the Laplace variable $s = i\omega$. The characteristic polynomial of the Green's function denominator transforms into a 4th-degree polynomial with real coefficients, $P(s) = \sum\limits_{k=0}^4 a_{4-k} s^k$. Expanding the determinant of the effective system yields the explicit coefficients: $a_0 = 1$, $a_1 = \eta$, $a_2 = g^2 + \frac{\eta^2}{4} + \frac{\eta\Gamma}{2}$, $a_3 = \eta (g^2 + \frac{\eta\Gamma}{4})$, and $a_4 = \frac{\eta^2}{4} (g^2 + \frac{\Gamma^2}{4})$. 

Evaluating the integral for a constant numerator $b_0 = \frac{1}{16}\eta^4 \Gamma^2 g^2$ yields a closed expression for the current:
\begin{equation}
J_{\text{max}}^{(N=2)} = \frac{2 \eta \Gamma g^2 (\eta + \Gamma)}{(4g^2 + \Gamma^2)(4g^2 + \eta^2)}
\end{equation}
This reveals analytically that coherent transport is not maximized in the structureless limit. By imposing the stationary condition $\partial_\eta J_{\text{max}}^{(N=2)} = 0$, we find that the integrated current admits a global analytical maximum at a finite memory timescale:
\begin{equation}
\eta_{opt}^{J} = \frac{2g}{\Gamma} \left( 2g + \sqrt{4g^2 + \Gamma^2} \right)
\end{equation}

For intermediate coupling regimes, this finite optimal memory actively boosts the coherent current above the standard Markovian bounds. This formula coincides with the numerically observed peaks in mean current, however, note these do not exactly coincide with those of the Fano factor. Trying the same method for minimizing  $F \approx 1 - \frac{\int \mathcal{T}^2 d\omega}{\int \mathcal{T} d\omega}$ is intractable as evaluating the component $\int \mathcal{T}^2 d\omega$ squares the rational denominator, thus mapping the integration into the $I_8$ James-Nichols-Phillips class.

To understand the origin of the Fano resonance, we must analyze the shape of the dispersive response before energy integration. As discussed above, the shape of the Green's function is dictated by the determinant in the denominator, which is a fourth-degree polynomial in $\omega$. Its roots are the complex eigenvalues of the non-Hermitian $H_{\text{eff}}$, denoted as $\tilde{\omega}_j = \Omega_j - i\gamma_j / 2$. Physically, the real part ($\Omega_j$) corresponds to the hybridized resonant frequency of the mode \cite{datta2005quantum}, and the imaginary part ($\gamma_j$) is its effective dissipation, fundamentally dictated by $\eta$ in this system.

By factoring the polynomial we can rewrite the Green's function as a sum of four independent linear fractions with some complex weights $A_j$:
\begin{equation}
\mathcal{G}^r_{14}(\omega) = \sum_{j=1}^{4} \frac{A_j}{\omega - \Omega_j + i \gamma_j / 2}
\end{equation}

According to the Kubo formula \cite{kubo1957statistical}, the susceptibility of a quantum system is given by the retarded Green's function of the corresponding operators. Therefore, the decomposition of $\mathcal{G}^r_{14}(\omega)$ into a sum of simple poles directly expresses the dynamical susceptibility of the system. The resulting equation is identical to the complex susceptibility $\chi(\omega)$ of a classical damped driven harmonic oscillator (DHO) \cite{Lorentz}. The system therefore acts as a superposition of damped resonant modes, where each pole at $\Omega_j-i\gamma_j/2$ represents a collective excitation with resonance frequency $\Omega_j$ and damping rate $\gamma_j$, while the residue $A_j$ determines its weight in the response.

The isolated contribution of each pole to the transmission probability, $|\frac{A_j}{ (\omega - \Omega_j + i\gamma_j/2)}|^2$ generates a Lorentzian lineshape. These functions are positive and symmetric with respect to their resonance frequency $\Omega_j$. We conclude that the non-monotonic behavior cannot originate from the direct transmission terms.

The dispersive minimum originates from interference cross-terms between different pole contributions. In the high-bias regime ($eV\gg T$), the zero-frequency noise gives
$F \approx 1-\int \mathcal{T}^2 d\omega/\int \mathcal{T}d\omega$.
A pronounced dip in $F$ therefore requires a transmission profile that remains close to unity over the active transport window. When the internal and external scales are detuned ($g\neq g_{\rm opt}$), the pole overlaps are reduced, the interference terms are suppressed by energy integration, and the Fano factor becomes monotonic. Although this overlap condition can be reached for small $g$ by tuning $\eta$, optimal transport additionally requires the Markovian impedance-matching condition discussed in Appendix~\ref{ap:crit}.

We can evaluate the integration of a generic cross-term between poles $j$ and $k$ over a wide energy band using the residue theorem to ensure the previous analytical result is consistent with the interference mechanism. In the high-bias regime:

\begin{equation}
\int_{-\infty}^{\infty} \left[\frac{A_j}{\omega - \Omega_j + i\gamma_j/2} \frac{A_k^*}{\omega - \Omega_k - i\gamma_k/2} + \text{c.c.} \right] d\omega
\end{equation}

Closing the contour in the upper half-plane captures only the pole at $\omega = \Omega_k + i\gamma_k/2$. Defining the energy splitting $\Delta\Omega = \Omega_j-\Omega_k$ and the average dissipation $\bar{\gamma}=(\gamma_j+\gamma_k)/2$, the residue evaluation yields the integrated interference contribution $I_{jk}^{\rm cross}$:
\begin{equation}
\label{eq:interference}
I_{jk}^{\rm cross} = 4\pi \frac{\text{Re}(A_j A_k^*) \bar{\gamma} + \text{Im}(A_j A_k^*) \Delta\Omega}{\Delta\Omega^2 + \bar{\gamma}^2}
\end{equation}
This expression shows that the cross-term remains sizable only when the linewidths are not small compared with the inter-pole splitting and when the residues have a favorable relative phase.

\subsection{Generalization to \textit{N}-site chains}
\label{ap:ndot}
We can extend the analysis from Appendix~\ref{ap:dqd} to a symmetrical linear chain of $N$ quantum dots coupled to the same two auxiliary memory leads. The total system therefore consists of $d=N+2$ sites. The arguments here are a direct generalization of the previous ones.

The effective non-Hermitian Hamiltonian $H_{\text{eff}}$ of dimension $d \times d$ is once again tridiagonal, containing the internal hoppings $g$ and the boundary couplings $\lambda$. The transmission function depends strictly on $\mathcal{G}_{1,d}^r(\omega)$.

Using Cramer's rule, the element $(1,d)$ of $G(\omega) = (\omega I - H_{\text{eff}})^{-1}$ is given by:
\begin{equation}
    \mathcal{G}_{1,d}^r(\omega) = \frac{(-1)^{1+d} \det(M_{d,1})}{\det(\omega I - H_{\text{eff}})}.
\end{equation}

For a linear chain with nearest-neighbor couplings, the minor matrix $M_{d,1}$ (obtained by removing the first column and last row) is lower triangular. Its determinant is simply the product of the off-diagonal coupling elements:
\begin{equation}
    \det(M_{d,1}) = \prod_{i=1}^{N+1} (-t_i) = (-\lambda)(-g)^{N-1}(-\lambda).
\end{equation}
The numerator is once more energy-independent. The denominator is the characteristic polynomial of degree $d$. The Green's function therefore admits a partial fraction decomposition:
\begin{equation}
\label{equation:indep}
    \mathcal{G}_{1,d}^r(\omega) = \sum_{j=1}^{N+2} \frac{A_j}{\omega - \Omega_j + i\gamma_j/2}.
\end{equation}

\begin{figure}[t]
    \centering
    \includegraphics[width=0.5\linewidth]{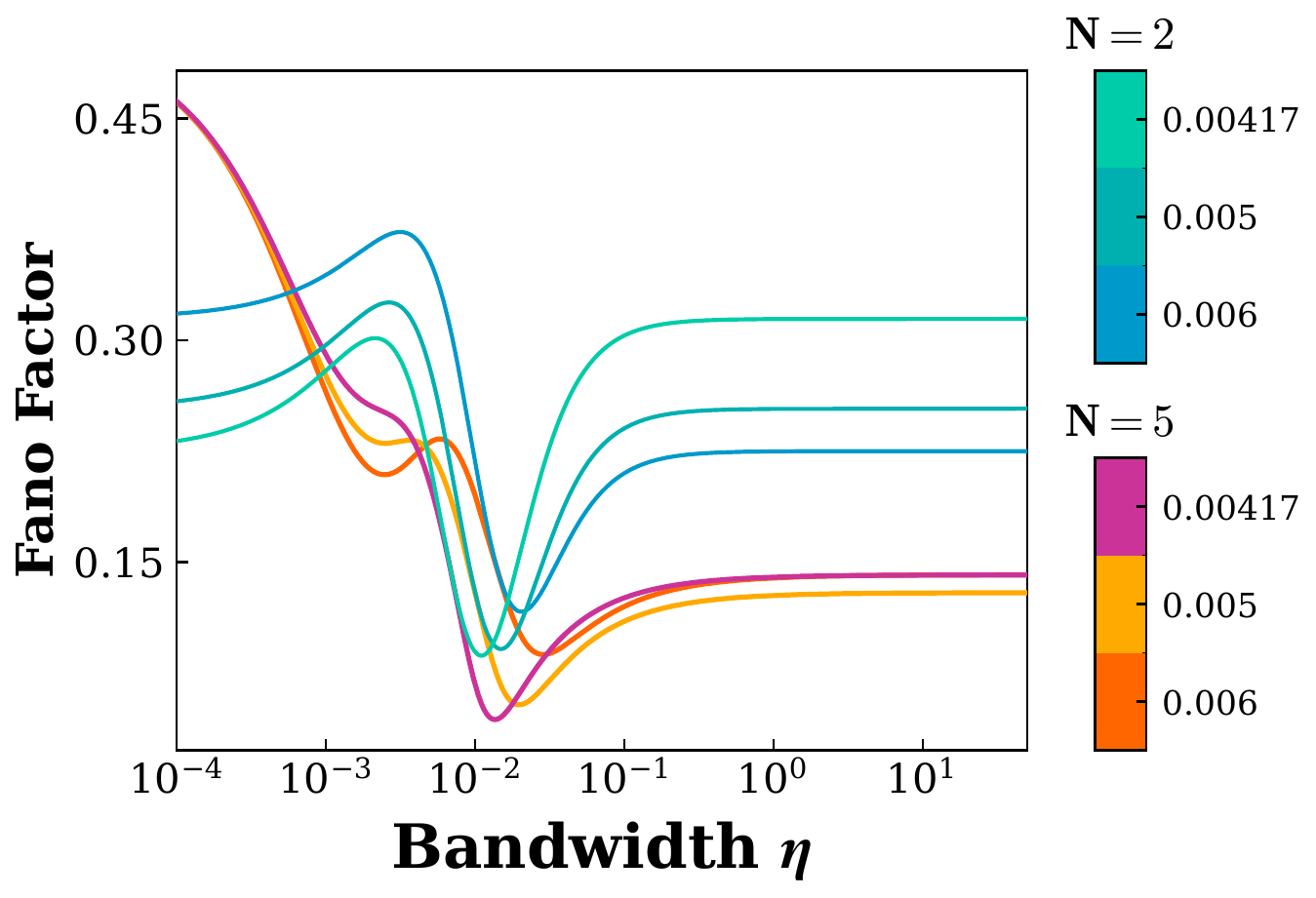}
    \parbox{\textwidth}{
    \caption{\justifying The current Fano factor $F$ is plotted with respect to bath bandwidth $\eta$ for even and odd chain lengths, $N=2$ and $N=5$, and several couplings $g$. The finite-memory minimum remains visible, but its position and depth depend on the parity of the chain. Parameters: $\Gamma_\alpha=0.01$, $T=1$, $eV/T=12$, $\varepsilon_{L_\alpha}=\varepsilon=1$.
    \label{fig:parity}}}
\end{figure}

The rest of the derivation follows the DQD case. Thus, dispersive memory-dependent features in $N$-dot chains arise from the same interference mechanism despite parity effects playing a role due to zero modes, this is shown in Fig~\ref{fig:parity}.

\section{Optimal Memory Scaling}
\label{ap:scale}

We now derive the conditions for the optimal memory rate $\eta_{\text{opt}}$ in symmetric chains of length $N$. Asymmetric configurations require numerical optimization~\cite{ehrlich2021broadband}. For $N>2$, the chain contains internal sites that are not directly coupled to the reservoirs; this makes the Markovian impedance-matching condition discussed in Appendix~\ref{ap:crit} necessary for optimal transport.

\subsection{The long-chain limit}
\label{ap:Nchainscale}
In the structureless Markovian limit, the boundary acts as a frequency-independent sink with decay rate $\Gamma/2$, leading to the critical-damping condition derived above. However, when introducing a Lorentzian bath the spectral density becomes frequency dependent:

\begin{equation}
\mathcal{J}_{eff}(\omega) = \frac{\lambda^2 \eta}{\omega^2 + (\eta/2)^2} 
\end{equation}

To maximize the mean current while simultaneously suppressing zero-frequency shot noise, the transmission function must approach a ripple-free profile. In classical networks, the optimal solution to this problem is the maximally flat Butterworth filter \cite{pozar2011microwave}. The power loss ratio $P_{LR}$ for such a filter is defined as:
$$
P_{LR} = 1 + k^2 \left(\frac{\omega}{\omega_c}\right)^{2N},
$$
where $N$ is the filter order and $\omega_c$ is the cutoff frequency. Mapping the classical transmission coefficient $|\mathcal{T}|^2 = 1/P_{LR}$ to our quantum system dictates that the ideal transmission probability must take the form:
$$
\mathcal{T}(\omega) = \frac{1}{1 + (\frac{\omega}{\omega_c})^{2N}}
$$
The defining mathematical characteristic of this Butterworth polynomial is that its first $2N-1$ derivatives with respect to $\omega$ strictly vanish at the band center ($\omega = 0$), creating a perfectly flat transmission plateau. In quantum transport, zero-frequency shot noise is governed by the integral $\int \mathcal{T}(\omega)(1-\mathcal{T}(\omega)) d\omega$. Consequently, any dispersive dip or internal ripple within the transport window ($\mathcal{T} < 1$) injects partition noise. By enforcing a maximally flat transmission geometry, these internal reflections are eliminated.

To optimize transmission, the structured bath must simultaneously satisfy two constraints. First, the Full Width at Half Maximum of the Lorentzian pseudomode must exactly encompass the internal bandwidth of the chain to prevent underdamping, which is \cite{ashcroft1976solid, kittel2004introduction}:
$$
\eta = W_{chain} = 4g
$$
Second, the structured bath must preserve a perfect impedance match at the Fermi level. Evaluating the effective spectral density of the reaction coordinate mapping at resonance yields:
$$
\mathcal{J}_{eff}(0) = \frac{\lambda^2 \eta}{(\eta/2)^2} = \frac{4\lambda^2}{\eta}
$$
Imposing the ideal reflectionless boundary condition $\mathcal{J}_{eff}(0) = 2g$ and substituting the macroscopic bandwidth requirement $\eta = 4g$, we extract the exact condition for the boundary coupling:
$$
\frac{4\lambda^2}{4g} = 2g \implies \lambda^2 = 2g^2 \implies \lambda = \sqrt{2}g
$$
Thus we have deduced a condition relating the pseudomode mapping with the chain parameters. Now, substituting this expression in Eq.~\ref{eq:rc} yields:
$$
\eta=\frac{8g^2}{\Gamma}
$$
This equation predicts the optimal parameters with high accuracy over a broad range centered around the Markovian critical-damping condition $\Gamma=2g$.

\subsection{The intermediate regime}
\label{ap:intermediate}

We can also obtain analytical estimates for the intermediate cases $N=3,4$, assuming the Markovian impedance-matching condition. For $N=3$, the system is already distinct from the DQD but remains far from the continuum-band limit. The system possesses a central protected dot. The tight-binding spectrum, $E_k = 2g \cos\left(\frac{k\pi}{N+1}\right)$ features a zero-energy mode ($E_{k=2} = 0$) fixed by the symmetry of the odd chain. Transport near the Fermi level is fundamentally mediated by the energetic splitting between this mode and its nearest neighbors. The relevant energy scale is therefore the bare spectral gap:
\begin{equation}
\Delta\Omega^{(N=3)} = 2g \cos{\left(\frac{\pi}{4}\right)} = \sqrt{2}g \approx 1.414g
\end{equation}

By virtue of the invariance of the trace of the non-Hermitian Hamiltonian, the sum of all decay rates strictly equals $2\eta$, yielding an average decay rate per mode of $\bar{\gamma} = \frac{2\eta}{N+2}$. To optimize transport, the mean decay rate of the hybridized collective modes of the extended system must match the gaps between modes. Equating $\frac{2\eta}{5} = \sqrt{2}g$ yields:

\begin{equation}
    \eta_{M}^{(N=3)} = \frac{5\sqrt{2}}{2} g_M \approx 3.535g
\end{equation}

This prediction agrees with the numerical optimum, $\eta_M\simeq3.47g$, indicating that the central mode controls the coherent transport in this regime.

For $N=4$, the same matching argument gives a comparable estimate. In this case, the spectrum lacks a zero-energy mode, so the coherent transport near the Fermi level is dominated by the splitting between the two central states. The relevant internal gap is:
\begin{equation}
    \Delta\Omega^{(N=4)} = 2g \left[ \cos\left(\frac{2\pi}{5}\right) - \cos\left(\frac{3\pi}{5}\right) \right] = (\sqrt{5}-1)g \approx 1.236g
\end{equation}
Applying the uniform hybridization condition for this chain, the mean decay rate of the collective modes is $\bar{\gamma} = \frac{2\eta}{N+2} = \frac{\eta}{3}$. Equating $\frac{\eta}{3} = (\sqrt{5}-1)g$ yields the exact analytical scaling:
\begin{equation}
    \eta_{M}^{(N=4)} = 3(\sqrt{5}-1) g_M \approx 3.708g
\end{equation}

This prediction closely approximates the numerically observed optimum of $\eta_M \approx 3.8g_M$. This deviation marks the onset of the transition towards the macroscopic continuum where the separation between the internal sites and the boundaries limits the penetration depth of the dissipation, breaking the uniform equipartition approximation.

\subsection{Numerical results}
\label{ap:numerics}
Before detailing the numerical optimization of the precision bounds, we establish the algebraic constraints imposed by the energy scales of the system. In the high-bias regime, the Fano factor depends only on $g$, $\eta$, and $\Gamma$. If all energy scales are multiplied by a factor $\lambda$, so that $H\to\lambda H$, then the retarded Green's function at frequency $\lambda\omega$ satisfies
\begin{equation}
        \mathcal{G}(\lambda \omega) = (\lambda \omega - \lambda H)^{-1} = \frac{1}{\lambda} (\omega - H)^{-1} = \frac{1}{\lambda} \mathcal{G}(\omega).
\end{equation}
The broadening matrices scale proportionally, and thus the transmission function yields:
\begin{equation}
    \mathcal{T}(\lambda \omega) = \text{Tr}\left[ (\lambda \Gamma_L) \frac{1}{\lambda} \mathcal{G}(\omega) (\lambda \Gamma_R) \frac{1}{\lambda} \mathcal{G}^\dagger(\omega) \right] = \mathcal{T}(\omega).
\end{equation}

Consequently, the Fano factor becomes:
\begin{equation}
    F(\lambda g, \lambda \Gamma, \lambda \eta) = \frac{\int_{-\infty}^{\infty} \mathcal{T}(\lambda\omega) [1 - \mathcal{T}(\lambda\omega)] d(\lambda\omega)}{\int_{-\infty}^{\infty} \mathcal{T}(\lambda\omega) d(\lambda\omega)} = F(g, \Gamma, \eta),
\end{equation}
since the differential $d(\lambda\omega) = \lambda d\omega$ extracts a global factor $\lambda$ that cancels out. By choosing $\lambda = 1/\Gamma$, we exploit this scale invariance to show that the Fano factor depends on the ratios of these parameters:
\begin{equation}
    F(g, \Gamma, \eta) = F\left(\frac{g}{\Gamma}, 1, \frac{\eta}{\Gamma}\right).
\end{equation}

This implies the optimum must lie inside a linear manifold, as the invariance of the transmission function restricts the degrees of freedom of the system.

We distinguish two optimization protocols: the global optimum $(g_G,\eta_G)$ and the two-step optimum $(g_M,\eta_M)$. In both cases, the long-chain memory scale follows $\eta=8g^2/\Gamma$; the difference lies in the value of $g$. We define $g_M$ as the one satisfying Markovian optimality, and $g_G$ as the global optimal value when minimizing in the space $(g,\eta)$. The difference between the optimal couplings is further discussed in Appendix~\ref{ap:transmission}.

\subsubsection{The two-step optimality}
\label{ap:twostep}
We check the consistency of the relations for $(g_M,\eta_M)$ in Table~\ref{tablaslopes}. Crucially, $\mathcal{M}$ remains constant in the long chain limit, and is stable against $\Gamma$ and $N$ changes.

\begin{table}[t]
    \centering
    \renewcommand{\arraystretch}{1.2}
    \begin{tabular}{ccccc}
        \hline\hline
        $N$ & $\frac{\eta_M}{g_M}$ Fano & $\frac{\eta_M}{g_M}$ $\mathcal{Q}_{QK}$ & $\bar{\mathcal{M}}$ & $\sigma_{\mathcal{M}}$ \\
        \hline
        2  & 3.6036 & 3.6090 & 0.7440 & $7.13 \times 10^{-4}$ \\
        3  & 3.4721 & 3.4695 & 0.7117 & $8.89 \times 10^{-4}$ \\
        4  & 3.8000 & 3.8157 & 0.7329 & $< 10^{-6}$ \\
        5  & 3.9497 & 3.9568 & 0.7411 & $8.98 \times 10^{-4}$ \\
        6  & 3.9770 & 3.9925 & 0.7439 & $6.38 \times 10^{-4}$ \\
        7  & 4.0000 & 3.9961 & 0.7445 & $< 10^{-6}$ \\
        8  & 4.0000 & 3.9961 & 0.7445 & $< 10^{-6}$ \\
        9  & 4.0000 & 3.9961 & 0.7445 & $< 10^{-6}$ \\
        10 & 4.0000 & 3.9961 & 0.7445 & $< 10^{-6}$ \\
        \hline\hline
    \end{tabular}
    \captionof{table}{ \justifying Slopes $\frac{\eta_M}{g_M}$ obtained through linear regression for Fano and $\mathcal{Q}_{QK}$ and memory-induced bottleneck results for several values of $N$, in particular mean value ($\bar{\mathcal{M}}$) and standard deviation ($\sigma_{\mathcal{M}}$). Results were obtained considering $\Gamma \in [0.01,0.1]$, $\Gamma_{1,2}=\Gamma$, $T=1$, $\frac{eV}{T}=20$, $\varepsilon_{\{D_\alpha,L_\alpha\}}=1$}
    \label{tablaslopes}
\end{table}

Although these pairs are not the true global minima, they are useful because they can be obtained through two consecutive one-dimensional optimizations. For $N>3$, the Markovian optimum is well approximated by $\Gamma=2g$. The resulting parameters remain close to the global optima while being substantially easier to compute.

\subsubsection{Variational search for the global thermokinetic optimum}
\label{ap:global}

Global minimization requires a slight deviation from $g_M$ to improve transport quality as discussed. The global minimum ($g_G,\eta_G$) must be found via an unconstrained two-dimensional minimization in the $(g, \eta)$ parameter space. By performing this numerical optimization, we allow the system to independently balance the level repulsion against the noise injection from the structured bath and the mean current. The results are shown in Table~\ref{tab:optimos_globales}.

\begin{table}[t]
    \centering
    \renewcommand{\arraystretch}{1.2}
    \begin{tabular}{ccccc}
        \hline\hline
        $N$ & $\frac{g_G}{\Gamma}$ & $\frac{\eta_G}{\Gamma}$ & $\frac{\eta_G}{g_G}$ & $\bar{\mathcal{M}}$ \\
        \hline
        2  & 0.440 & 1.200 & 2.729 & 0.6437 \\
        3  & 0.440 & 1.346 & 3.061 & 0.6739 \\
        4  & 0.432 & 1.407 & 3.256 & 0.6895 \\
        5  & 0.425 & 1.405 & 3.309 & 0.6934 \\
        6  & 0.421 & 1.407 & 3.342 & 0.6958 \\
        7  & 0.418 & 1.396 & 3.338 & 0.6954 \\
        8  & 0.418 & 1.393 & 3.336 & 0.6953 \\
        9  & 0.417 & 1.391 & 3.336 & 0.6952 \\
        10 & 0.417 & 1.390 & 3.335 & 0.6952 \\
        \hline\hline
    \end{tabular}
\captionof{table}{\justifying Optimal transport parameters scaling as a function of the number of quantum dots ($N$) in the chain. Notably, the optimization yields identical global minima coordinates for both the Fano factor and the QKUR bound within the numerical precision of the methods employed. The parameters are normalized against the coupling strength $\Gamma$ and the ratios are invariant across the studied values of $\Gamma$. $\bar{\mathcal{M}}$ represents the average memory-induced bottleneck measure, which exhibits negligible variance for any given $N$. Parameters: $0.01 \leq \Gamma \leq 0.1$, $T=1$, $\frac{eV}{T}=10$}.
\label{tab:optimos_globales}
\end{table}

\subsection{Memory-induced coupling deviations and the shape of the bath}
\label{ap:transmission}

The departure of the global optimal coupling $g_{G}$ from the Markovian impedance-matching condition $\Gamma = 2g$ can be understood as a trade-off between maximizing resonant transmission and filtering high-frequency shot noise. 

From the high-bias expression
$$
F = \frac{\int \mathcal{T}(1-\mathcal{T})}{ \int \mathcal{T}}
$$

\begin{figure}[b]
    \centering
    \includegraphics[ trim= .2cm 0.2cm 0.2cm 0.cm, clip,width=0.5\linewidth]{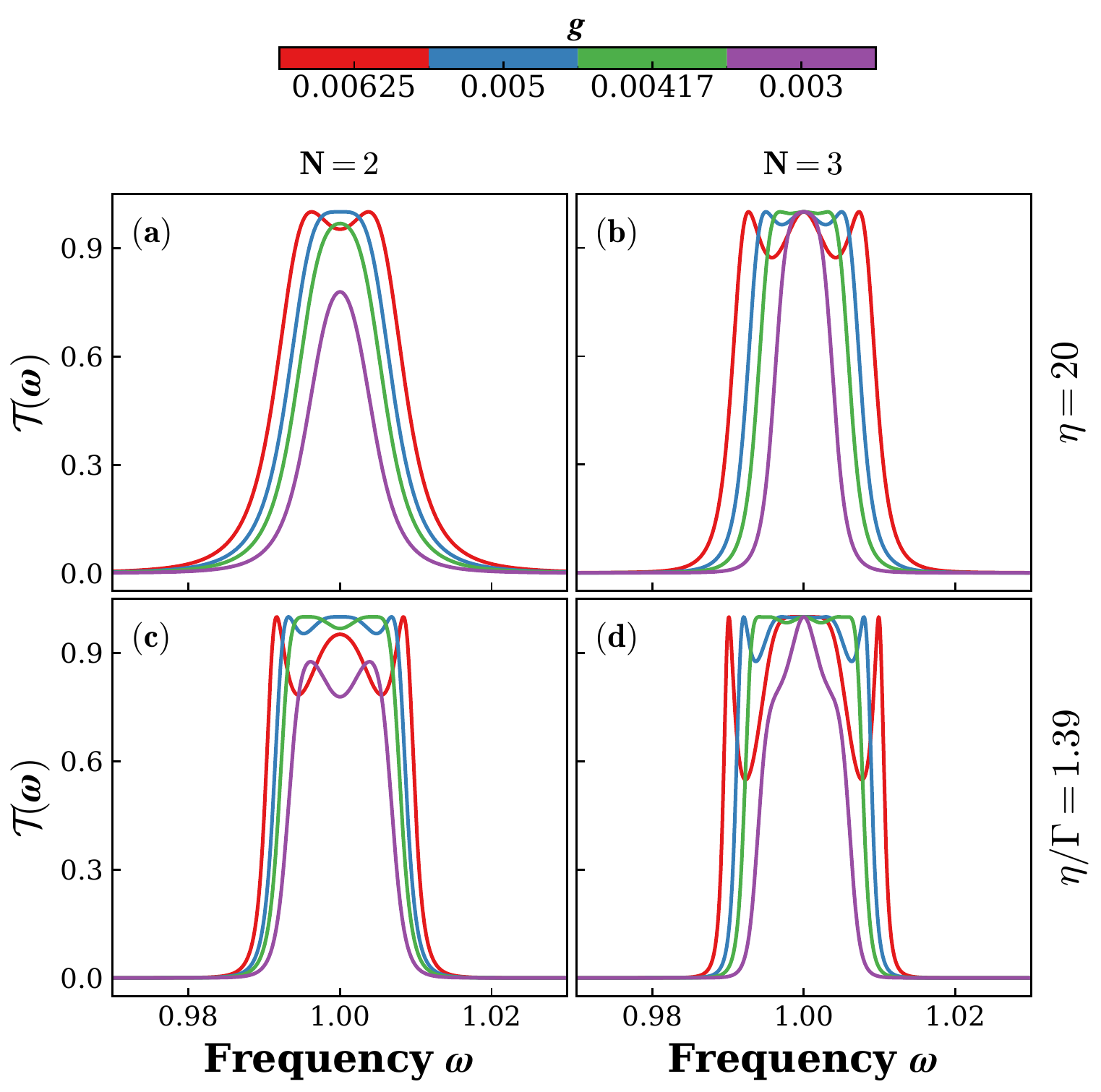}
    \parbox{\textwidth}{
    \caption{\justifying Transmission function near resonance in the Markovian regime, $\eta=20$, and in the optimal-memory regime, $\eta/\Gamma=1.417$. Parameters: $\Gamma_\alpha=0.01$, $T=1$, $eV/T=12$, $\varepsilon_{L_\alpha}=\varepsilon=1$.
    \label{transmission}}}
\end{figure}

The Fano factor is reduced by keeping $\mathcal{T}\approx1$ in the active transport window, suppressing $\mathcal{T}$ in the tails, or increasing the integrated transmission $\int\mathcal{T}d\omega$. Boxcar-like profiles are known to be favorable for transport optimization~\cite{ehrlich2021broadband}. In the present system, slightly less flat profiles can still be favored if they increase the mean current. Representative transmission functions are shown in Fig.~\ref{transmission}. Consistent with the critical-damping analysis, the Markovian impedance-matching condition gives optimal transmission at $\omega=0$.

For a Lorentzian bath, the effective dissipation decays away from the resonance center and Lamb shift effects are introduced. If the internal coupling $g$ is kept at its Markovian optimal value, the level repulsion pushes the side poles into the tails of the Lorentzian spectral density. In these off-resonant regions, the side modes experience reduced dissipation, becoming underdamped and sharply peaked. These narrowed side resonances fail to sufficiently overlap with the central mode, breaking the plateau and opening deep transmission valleys that inject substantial noise. To flatten the transmission window, the system is forced to reduce its internal coupling. This reduction pulls the side poles back toward the stronger dissipation region of the bath, allowing them to adequately broaden and merge once again. The DQD behaves differently because it lacks internal bulk modes. Its wide gap, $\Delta\Omega=2g$, cannot be bridged efficiently by the boundary dissipation. To maximize the integrated transmission, the DQD therefore accepts a deeper central valley and compensates through stronger coupling, which explains why its optimal Fano values are higher than those obtained for $N\ge3$.

\section{Analytical details for other bath structures}
\label{ap:other}

In the main text, we established that the memory-induced noise suppression mechanism is robust across different structured environments. Since baths with arbitrary spectral densities cannot generally be mapped exactly to a finite pseudomode expansion, their thermodynamic and transport observables are evaluated numerically using the NEGF formalism. However, the locations of the Fano noise minima can be analytically approximated by extending the impedance matching conditions of the Lorentzian model through the renormalization of the Lamb shift produced by other spectral structures. 
For the Lorentzian spectral density [Eq.~\eqref{eq:lorentzian_density_results}], the Lamb shift is
\begin{equation}
    \Delta_L(\omega_r, \eta_{\text{opt}}) =  \mathcal{P} \int \frac{d\omega'}{2\pi}\frac{\mathcal{J}_L(\omega')}{\omega - \omega'}= \frac{\Gamma \left(\frac{\eta_{\text{opt}}}{4}\right) \omega_r}{\omega_r^2 + \left(\frac{\eta_{\text{opt}}}{2}\right)^2}.
\end{equation}

Here we provide the derivations of the base bandwidths and the Lamb shifts necessary to evaluate the scaling ansatz Eq.~\eqref{eq:scaling_ansatz} for both the Newns and Gaussian spectral densities.

\subsection{The Newns Spectral Density}
\label{ap:newns}

The Newns bath models the spectral properties of a semi-infinite tight-binding chain of quantum dots. Its spectral density features a hard cutoff and is given by:
\begin{equation}
    \mathcal{J}_N(\omega) = \Gamma \sqrt{1 - \left(\frac{\omega - \omega_0}{\omega_c}\right)^2} \quad \text{for} \quad |\omega - \omega_0| \le \omega_c,
\end{equation}
where $\omega_c$ is the half-bandwidth and $\omega_0$ is the spectral center. We restrict our analysis to the region within the band, as dissipation strictly vanishes outside ($\mathcal{J}_N = 0$).

To evaluate the matching condition, we first determine the base cutoff $\omega_{\text{base}}$ that equates the local dissipation at the bare lateral poles $\omega_r = 2g \cos\left(\frac{\pi}{N+1}\right)$ to the target Lorentzian dissipation. Assuming $\omega_0 = 0$, algebraic inversion yields:
\begin{equation}
    \omega_c^{\text{base}} = \frac{\omega_r}{\sqrt{1 - (\mathcal{J}_{\text{target}}/\Gamma)^2}}.
\end{equation}

The real part of the bath self-energy (the Lamb shift) is defined through the Kramers-Kronig relation:
\begin{equation}
    \Delta_N(\omega) = \frac{1}{2\pi} \mathcal{P} \int_{-\omega_c}^{\omega_c} \frac{\mathcal{J}_N(\omega')}{\omega - \omega'} d\omega'.
\end{equation}
Evaluating this inside the band ($|\omega| < \omega_c$) yields a linear dependence on the energy:
\begin{equation}
    \Delta_N(\omega, \omega_c) = \frac{\Gamma}{2} \left( \frac{\omega}{\omega_c} \right).
\end{equation}

Substituting $\omega = \omega_r$ and $\omega_c =\omega_c^{\text{base}}$ into this expression provides the displacement necessary to compute the renormalized optimal cutoff $\omega_c^{\text{opt}}$ via Eq.~\eqref{eq:scaling_ansatz}. As discussed in the main text, the predicted optimal cutoffs successfully track the deep memory minima, avoiding the destruction of the interference cross-terms that would occur if the repelled poles were pushed beyond the hard cutoff.

\subsection{The Gaussian Spectral Density}
\label{ap:gauss}

Unlike the Newns model, the Gaussian spectral density lacks a hard energy cutoff, extending over the entire continuous spectrum. It is defined as:
\begin{equation}
    \mathcal{J}_G(\omega) = \Gamma \exp\left[- \left(\frac{\omega - \omega_0}{\omega_c}\right)^2\right],
\end{equation}
where $\omega_c$ parameterizes the effective bandwidth (proportional to the standard deviation) and $\omega_0$ its center. 

By enforcing $\mathcal{J}_G(\omega_r, \omega_c^{\text{base}}) = \mathcal{J}_{\text{target}}$ for a symmetric bath centered at $\omega_0=0$, we can find the necessary bandwidth for the match:
\begin{equation}
   \omega_c^{\text{base}} = \frac{\omega_r}{\sqrt{\ln\left(\Gamma / \mathcal{J}_{\text{target}}\right)}}.
\end{equation}

The Lamb shift induced by the Gaussian takes the form:
\begin{equation}
    \Delta_G(\omega) = \frac{\Gamma}{2\pi} \mathcal{P} \int_{-\infty}^{\infty} \frac{e^{-(\omega'/\omega_c)^2}}{\omega - \omega'} d\omega'.
\end{equation}
This principal-value integral can be written in terms of the Dawson function $F_D(x)=e^{-x^2}\int_0^x e^{y^2}dy$, giving
\begin{equation}
    \Delta_G(\omega, \omega_c) = \frac{\Gamma}{\sqrt{\pi}} F_D\left(\frac{\omega}{\omega_c}\right).
\end{equation}
Evaluating this function at the base bandwidth $\omega_c =\omega_c^{\text{base}}$ allows for the extraction of the shifted parameter via Eq.~\eqref{eq:scaling_ansatz}.

\begin{figure}
    \centering
    \includegraphics[width=\linewidth]{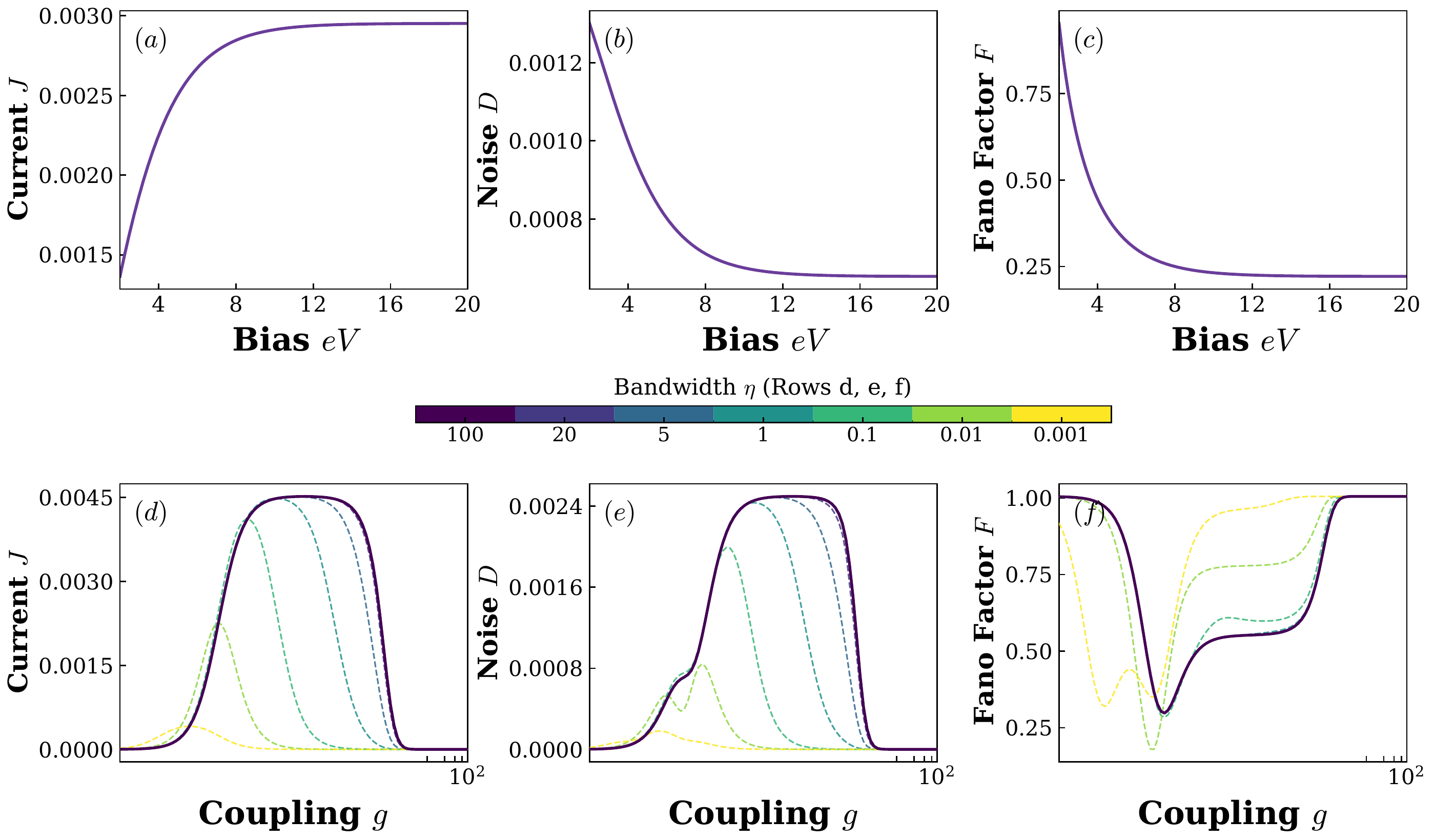} 
    \caption{\justifying (a--c) Markovian benchmark for $J$, $D$ and Fano Factor with $\eta=20$ and $g=0.006$ for different values of bias $eV$
    (d--f) Response to $g$ of $J$, $D$ and Fano Factor for different values of $\eta$ and $eV=6$. Parameters: $N=2$,$\Gamma_\alpha=0.01$, $T=1$,
    $\varepsilon_{L_\alpha}=\varepsilon=1$. Results are in excellent agreement with Ref.~\cite{markovian}.\label{fig:Markovian_benchmark}}
\end{figure}
\section{Markovian Limit Benchmark}
\label{appendixc}

We benchmark our NEGF calculations against the Markovian results of Ref.~\cite{markovian}. In Fig.~\ref{fig:Markovian_benchmark}(a--c), we set
$\eta=20$, which is already in the effectively flat-bath regime for the main parameter window considered here, and plot $J$, $D$, and $F$ as functions of the bias for $g=0.006$. The resulting curves reproduce the expected Markovian behavior.

As an additional check, Fig.~\ref{fig:Markovian_benchmark}(d--f) shows $J$, $D$, and $F$ as functions of $g$ for several values of $\eta$. The large-$\eta$ curves converge to the Markovian reference, while smaller $\eta$ values display the expected deviations due to finite reservoir memory. The small residual difference between $\eta=20$ and $\eta=100$ at
large $g$ indicates that stronger interdot coupling requires a larger bath broadening to reach the strict Markovian limit. This does not affect the main parameter regime, where $g<0.1$ and $\eta=20$ is sufficient for a Markovian behavior.

\end{document}